\documentclass[11pt,a4paper]{scrartcl}

\usepackage[T1]{fontenc}
\usepackage{lmodern}           
\usepackage{amsfonts, amsmath, amsthm, amssymb, dsfont}
\usepackage{graphicx}
\usepackage{epstopdf}
\usepackage{color}
\usepackage[breaklinks]{hyperref}

\usepackage{natbib}
\newcommand{\s}{\footnotesize}  %controll the font size of figure labels
\newcommand{\xs}{\scriptsize} %controll the font size of axis tics
\newcommand{\hl}[1]{\setlength{\fboxsep}{2.0pt}\colorbox{white}{#1}} %highlights figure labels with white background

\title{Spectral stability of nonlinear gravity waves in the atmosphere}
\author{Mark Schlutow\footnote{mark.schlutow@fu-berlin.de}, Erik Wahl\'en, Philipp Birken\\ \s Centre for Mathematical Sciences, Lund University, PO Box 118, 221 00 Lund, Sweden.}
\date{}

\begin{document}

%% Necessary!
\maketitle

\abstract{We apply spectral stability theory to investigate nonlinear gravity waves in the atmosphere. 
		These waves are determined by modulation equations that result from Wentzel-Kramers-Brillouin theory.		
		First, we establish that plane waves, which represent exact solutions to the inviscid Boussinesq equations, 
		are spectrally stable with respect to their nonlinear modulation equations
		under the same conditions as what is known as modulational stability from weakly nonlinear theory.
		In contrast to Boussinesq, the pseudo-incompressible regime does account for the altitudinal varying background density.
		Second, we show for the first time that upward-traveling wave fronts solving the inviscid modulation equations, 
		that compare to pseudo-incompressible theory,
		are unconditionally unstable.
		Both inviscid regimes turn out to be ill-posed as the spectra 
		allow for arbitrarily large instability growth rates.
		Third, a regularization is found by including dissipative effects. 
		The corresponding traveling wave solutions have localized amplitude 
		and blow up unconditionally by embedded eigenvalue instabilities
		but the instability growth rate is bounded from above.
		Additionally, all three types of nonlinear modulation equations
		are solved numerically to further investigate 
		and illustrate the nature of the analytic stability results.}

%%%%%%%%%%%%%%%%%%%%%%%%%%%%%%%%%%%%%%%%%%%%%%%%%%%%%%%%%%%%%%%%%%%%%
% MAIN BODY OF PAPER
%%%%%%%%%%%%%%%%%%%%%%%%%%%%%%%%%%%%%%%%%%%%%%%%%%%%%%%%%%%%%%%%%%%%%
%
	\section{Introduction}
	
%--------------------------------------------------------------------------------------
%%%%%%%%% Motivation for nonlinear gravity waves in the atmosphere
	Gravity waves have a significant impact on the dynamics of the earth's atmosphere.
	Since the resolution and the upper boundary of numerical weather and climate models steadily increase,
	a better understanding of these waves becomes necessary in order to 
	construct more precise subgrid-scale parametrizations \citep{Fritts2003}.
	Most gravity waves are excited in the lower atmosphere.
	However, they may propagate deep into the higher atmospheric layers above the stratopause.
	In these regions amplitudes have become so large, due to the thin background air, 
	that linear theory is not applicable anymore.
	
%%%%%%%%% weak-asymptotic nonlinear WKB theory and difference to weakly nonlinear theory
	Pioneering work on nonlinear gravity waves was accomplished by \citet{Bretherton1966} and \citet{Grimshaw1972}.
	These authors applied Wentzel-Kramers-Brillouin theory 
	to find leading-order asymptotic equations to the Euler equations.
	These asymptotic equations may be called modulation equations as they describe the evolution 
	of the wave characteristics like amplitude, frequency and wave number \citep{Whitham1974a}.

%%%%%%%%% Short overwiew about spectral stability analysis
	The stability of gravity waves is of particular interest 
	for modelers of subgrid-scale parametrizations to predict the wave's breaking height.
	Spectral stability theory is an especially useful approach 
	to study nonlinear waves \citep{Sandstede2000,Sandstede2002,Kapitula2013}.
	The key idea is to linearize the equations around a given wave solution
	which translates the problem of stability to finding the spectrum of a differential operator $\mathcal{L}$.
	The spectrum consists of all complex $\lambda$ for which the operator $\mathcal{L}-\lambda$ is not invertible.
	
%%%%%%%%% Results and Outline of the paper 
	The paper is structured as follows.
	We consider two-dimensional gravity waves being horizontally homogeneous in an isothermal atmosphere.
%%%%%%%%%%%%%%%%% inviscid Boussinesq
	Our analysis will start with the inviscid Boussinesq regime in section \ref{sec:bouss}. 
	Here, plane waves do not only fulfill the nonlinear modulation equations analytically 
	but also the Boussinesq equations themselves.
	From weakly nonlinear theory it is established in the literature that the plane wave is prone to
	modulational instabilities.
	We apply spectral stability analysis on the nonlinear modulation equations
	and find the very same criterion for modulational instability.
	The nature of the instability is also examined numerically. 

%%%%%%%%%%%%%%%%% inviscid Pseudo-incompressible
	The Boussinesq equations are limited in the sense that they do not cover anelastic growth of the amplitude.
	When a gravity wave propagates upwards its amplitude may increase due to the decreasing background density.
	This phenomenon influences the breaking height as the stability depends sensitively on amplitude.
	Therefore, we continue by investigating an extended set of modulation equations in section \ref{sec:psinc}
	which agrees with the inviscid pseudo-incompressible regime \citep{Durran1989,Achatz2010}.
	Here, the background density is an explicit function of height.
	These modulation equations possess upward-traveling wave solutions that take the form of a front
	where the envelope velocity is always greater than the derivative of the frequency with respect to the vertical wave number, 
	what one may call the linear group velocity. 
	That those two velocities are not the same, is a consequence of the nonlinearity.
	The spectral stability analysis reveals that the traveling wave fronts are unconditionally unstable.
	The dynamics of the instabilities will be illustrated numerically.
	
%%%%%%%%%%%%%%%%% dissipative Grimshaw
	The spectra of both inviscid regimes allow instabilities with arbitrarily large growth rates 
	which is evidence for ill-posedness \citep{Fetecau2011}. 
	In section \ref{sec:grimshaw},
	a regularization will be found by including dissipation 
	to the pseudo-incompressible modulation equations
	in accordance with the compressible Navier-Stokes equations.
	These modulation equations exhibit both upward and downward-traveling wave solutions
	where the amplitude is of finite extent making them wave packets.
	In terms of the spectral stability analysis we find out that 
	the wave packets are also unconditionally unstable.
	The waves destabilize by a superposition of exponentially growing eigenfunctions. 
	In contrast to the inviscid waves, a maximum instability growth rate can be computed explicitly.
	We will solve the modulation equations numerically for the wave packets 
	to demonstrate the evolution of these instabilities.

%--------------------------------------------------------------------------------------

	\section{Inviscid Boussinesq plane waves}
	\label{sec:bouss}

	We start our investigations with the modulation equations for gravity waves
	in the Boussinesq approximation. 
	Two spatial dimensions $(X,Z)$, one horizontal and one vertical, are assumed and the wave shall be homogeneous in $X$.
	For the derivation of these equations see, e.g., \citet{Muraschko2015}.
	We may give this coupled set of non-dimensionalized nonlinear equations here in conservative flux form
	\begin{align}
		\label{eq:boussi_mod}	  
    	&\partial_Tk_z+\partial_Z\bigl(\hat{\omega}(k_z)+K_xu\bigr)=0,\notag\\
	  	&\partial_T\mathcal{A}+\partial_Z\bigl(\hat{\omega}'(k_z)\mathcal{A}\bigr)=0,\notag\\
    	\rho&\partial_Tu+\partial_Z\bigl(\hat{\omega}'(k_z)K_x\mathcal{A}\bigr)=0.
	\end{align}
	The prognostic variables $k_z$, $\mathcal{A}$ and $u$ 
	denote the vertical wavenumber, the wave action density and the mean-flow horizontal wind, respectively.
	Their domain in $Z$ is the real line.
	The horizontal wavenumber $K_x$ is constant and without loss of generality positive, hereinafter. 
	Two additional positive constant parameters, the Brunt-V\"ais\"al\"a frequency $N$ and the background density $\rho$, 
	account for the ambient atmosphere which is assumed to be uniformly stratified.	
	The intrinsic frequency is defined throughout this article by the dispersion relation for non-hydrostic gravity waves 
	as a function of $k_z$, so
	\begin{align}
		\label{eq:dispersion}
		\hat{\omega}(k_z)=\frac{NK_x}{\sqrt{K_x^2+k_z^2}}
	\end{align}
	and its derivative
	\begin{align}
		\hat{\omega}'(k_z)=-\frac{NK_xk_z}{\sqrt{K_x^2+k_z^2}^3}
	\end{align}
	which denotes the linear group velocity.
	Throughout this article primes denote the derivative with respect to the vertical wave number.
	
	The first equation of \eqref{eq:boussi_mod} describes the conservation of the wave's troughs and crests.
	Its flux is exactly the extrinsic frequency, i.e. the Doppler-shifted intrinsic frequency.
	The second equation governs the conservation of wave action density 
	being the ratio of wave energy density and the intrinsic frequency.
	The third equation accounts for the acceleration of the mean flow 
	which is related to the Stokes drift known from water waves.
	
	The sign of $\hat{\omega}$ is ambiguous, but can be chosen without loss of generality to be positive
	due to the symmetry of the problem. 
	Consequently, the wave action density has to be strictly positive
	since $\hat{\omega}\mathcal{A}$ represents the wave energy density.
	
	\subsection{Plane wave solution}
	For simplification we introduce the ``specific'' wave action density $a$ by
	\begin{align}
		\mathcal{A}(Z,T)=\rho a(Z,T)
	\end{align}
	and rewrite system \eqref{eq:boussi_mod} in vector form
	\begin{align}
		\label{eq:boussi_mod_vec}	  		
		\partial_Ty+\partial_ZF(y)=0
	\end{align}
	for our solution vector $y=(k_z,\, a,\, u)^\mathrm{T}$ 
	with the flux function
	\begin{align}
		F(y)=\begin{pmatrix}
			\hat{\omega}(k_z)+K_xu\\
			\hat{\omega}'(k_z)a\\
			\hat{\omega}'(k_z)K_xa
		\end{pmatrix}.
	\end{align}
	Equation \eqref{eq:boussi_mod_vec} has a stationary solution $Y=(K_z,\, A,\, U)^\mathrm{T}=const.$
	that satisfies 
	\begin{align}
		\label{eq:bouss_plane_wave}
		\partial_TY=0\quad\text{and}\quad\partial_ZF(Y)=0
	\end{align}
	as $F$ is autonomous which corresponds to a monochromatic plane wave.
	It is well known that plane waves are solutions 
	to the fully nonlinear Boussinesq equations \citep{Mied1976a}.

	\subsection{Spectral stability of the plane waves}
	\label{sec:bouss_stab}
	
	In order to investigate the stability of the plane waves $Y$, 
	we perturb \eqref{eq:boussi_mod_vec} according to $Y+y$ 
	and linearize around the stationary solution,
	which yields
	\begin{align}
		\label{eq:boussi_mod_vec_lin}	  		
		\partial_Ty+\mathrm{D}F(Y)\partial_Zy=0
	\end{align}
	with the Jacobian
	\begin{align}
		\mathrm{D}F(Y)=\begin{pmatrix}
			\hat\omega'(K_z)&   0&              K_x\\
    		A\hat{\omega}''(K_z)&    \hat\omega'(K_z)&    0\\
    		K_xA\hat{\omega}''(K_z)& K_x\hat\omega'(K_z)& 0
		\end{pmatrix}.
	\end{align}
	Since the Jacobian of the flux does not depend on time,
	\eqref{eq:boussi_mod_vec_lin} must have a solution of the form
	\begin{align}
		\label{eq:eigenval_ansatz}
		y(Z,T)=y(Z)e^{\lambda T}
	\end{align}
	that gives us an eigenvalue problem
	\begin{align}
	\label{eq:boussi_evp}
		(\mathcal{L}_Y-\lambda)y=0
	\end{align}
	involving the differential matrix operator 
	\begin{align}
		\mathcal{L}_Y=-\mathrm{D}F(Y)\partial_Z.
	\end{align}		
	
	As $\mathcal{L}_Y$ is a differential operator 
	and therefore operates on an infinite-dimensional space, 
	in general one cannot expect to obtain only discrete eigenvalues, 
	$\lambda\in\mathbb{C}$,
	like one would get for finite-dimensional matrices.
	Instead $\mathcal{L}_Y$ generates a spectrum 
	which is the set of $\lambda$'s such that 
	the operator $\mathcal{L}_Y-\lambda$ is not invertible.
	The stability problem can be transfered to finding the spectrum of the operator $\mathcal{L}_Y$.
	We say the solution $Y$ is spectrally stable 
	if the spectrum of $\mathcal{L}_Y$ is 
	contained in the left hand side of the complex plane,
	so for every $\lambda$ in the spectrum $\Re(\lambda)\leq 0$ must hold.
	We can motivate this definition when we consider \eqref{eq:eigenval_ansatz}.
	If there is a point $\lambda$ in the spectrum with positive real part, 
	then the corresponding eigenfunction for the perturbation grows exponentially in time
	with the real part of $\lambda$ being the instability growth rate and the solution blows up.
	For negative or vanishing real part 
	the eigenfunction decays or remains bounded in time, 
	respectively, such that the solution is stable.
	
\begin{figure}[ht]
		\centering
		\input{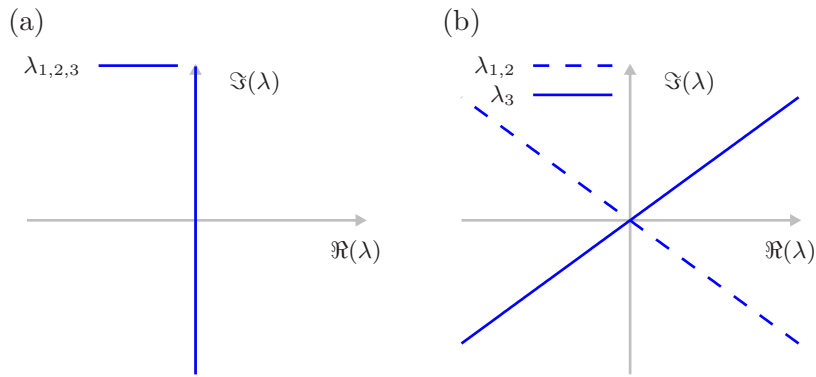}
		\caption{Spectra of $\mathcal{L}_Y$ for the inviscid plane Boussinesq wave. 
		Panel (a): Stable wave where $\hat{\omega}''(K_z)\geq 0$.
		Panel (b): Unstable wave where $\hat{\omega}''(K_z)<0$.}
		\label{fig:bouss_spec}
	\end{figure}

	We consider $\mathcal{L}_Y$ as a closed, 
	densely defined operator on $L^2(\mathbb{R},\mathbb{R}^3)$,
	the space of vector valued square integrable functions on the real line
	equipped with the norm
	\begin{align}
		\|y\|_{L^2}=\left(\int_\mathbb{R}y^\mathrm{T}y\,\mathrm{d}Z\right)^{1/2}.
	\end{align}	 
	The eigenvalue problem \eqref{eq:boussi_evp} can be reformulated as a linear
	ordinary differential equation (ODE)
	\begin{align}
		\label{eq:bouss_ode}
		\partial_Zy=B(\lambda)y
	\end{align}	 
	where $B(\lambda)=-\mathrm{D}F^{-1}(Y)\lambda$ is a spatially constant matrix. 
	The ODE \eqref{eq:bouss_ode} has a general solution in terms of the Fourier transform
	\begin{align}
		\label{eq:bouss_fourier}
		y(Z)=\int_\mathbb{R}\tilde{y}(\mu)e^{i\mu Z}\,\mathrm{d}\mu
	\end{align}
	which translates the ODE into an algebraic equation
	\begin{align}
		\bigl(B(\lambda)-i\mu\bigr)\tilde{y}=0
	\end{align}
	having nontrivial solution only if $\det(B(\lambda)-i\mu)=0$.
	This constraint gives us a parametrization for the spectrum of $\mathcal{L}_Y$ 
	in the complex plane by three different zeros
	\begin{align}
		\label{eq:bouss_spectrum_1}
		\lambda_1(\mu)&=~0,\\
		\label{eq:bouss_spectrum_2}
		\lambda_2(\mu)&=-i\mu\hat{\omega}'(K_z)+ i\mu\sqrt{K_x^2A\hat{\omega}''(K_z)},\\
		\label{eq:bouss_spectrum_3}
		\lambda_3(\mu)&=-i\mu\hat{\omega}'(K_z)- i\mu\sqrt{K_x^2A\hat{\omega}''(K_z)}.
	\end{align}
	The spectrum in the complex plane is depicted in figure \ref{fig:bouss_spec}. 
	It consists of straight lines going through the origin.
	As long as the discriminant in \eqref{eq:bouss_spectrum_2} and \eqref{eq:bouss_spectrum_3} 
	is positive the spectrum is the imaginary line (panel a)
	and we can conclude that the plane wave is spectrally stable. 
	If however the discriminant is negative, the spectrum splits into 
	two straight lines that extend into the right half of the complex plane 
	(panel b) and the wave becomes spectrally unstable.
	This can only happen when $\hat{\omega}''(K_z)$ becomes negative,
	which occurs if
	\begin{align}
		\label{eq:bouss_crit}
		\sqrt{2}|K_z|<K_x.
	\end{align}
	This result is well known and commonly referred 
	to as modulational instability.
	\citet{Whitham1974a}, \citet{Grimshaw1977} and \citet{Sutherland2001} derived this instability criterion
	from weakly nonlinear theory where the amplitude of the wave is used as the expansion parameter.
	Therefore, the coupling to the mean flow is a higher order effect.
	However, the modulation equations from the WKB theory, that are presented here, 
	turn out to be nonlinear in the leading order
	as the wave-mean-flow interaction and the Doppler shifting of the frequency 
	appear in the coupled set of leading order modulation equations.
	
	With the stability analysis of this section we know under what circumstances a plane wave becomes modulationally unstable,
	but how does the perturbation of an unstable wave evolve in time?

	\subsection{Numerical simulation of the plane waves}

	We explore the nature of the modulational instability further 
	by numerical computations of the evolution of the unstable plane wave.
	For this endeavor a finite-volume scheme was implemented to solve \eqref{eq:boussi_mod}.
	A precise description of the algorithm is given in appendix A.
	\begin{figure}[ht]
		\centering
		\input{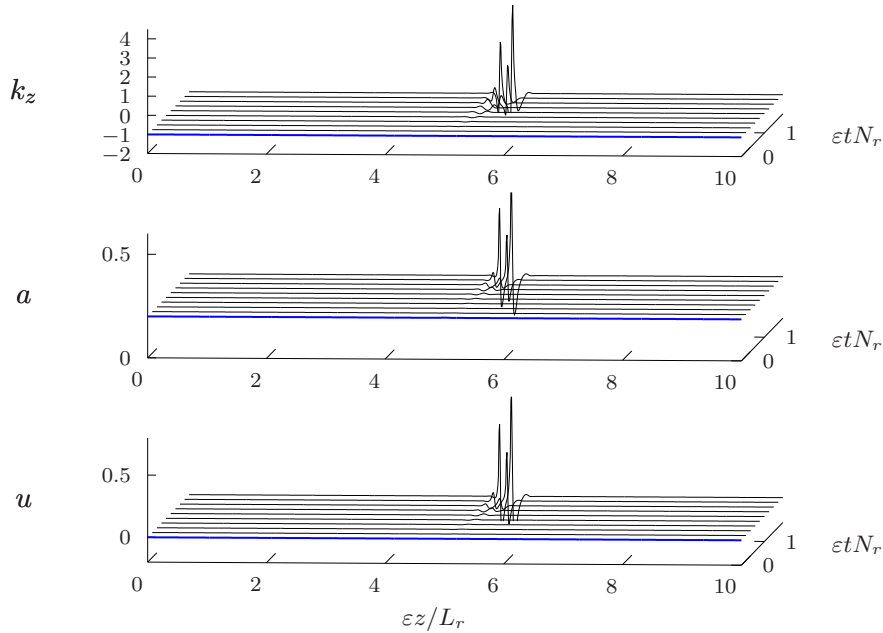}
		\caption{Numerical simulation of the modulationally unstable plane inviscid Boussinesq wave 
		initially perturbed by a small peak in the middle of the domain showing absolute instability.}
		\label{fig:boussi_3d}
	\end{figure}
	The computational results are plotted in figure \ref{fig:boussi_3d}
	for an initially unstable wave according 
	to the modulational stability criterion of \eqref{eq:bouss_crit}.
	The axes are labeled by the dimensional height $z$ and time $t$.
	The length scale is fixed by a reference wave length $L_r$ and the time scale 
	is given by a reference Brunt-V\"ais\"al\"a frequency $N_r$.
	The small scale separation parameter $\varepsilon$, being the WKB-expansion parameter, 
	measures the ratio between $L_r$ and the length scale on which the wave's envelope varies.	
	To seed the instability, the initial conditions are perturbed 
	by a Gaussian peak in the middle of the domain
	of the order of magnitude $10^{-5}$. 
	The parameters are set to $N=1.2$ and $K_x=1.8$.
	We observe that the perturbation, that excites instability modes, 
	blows up quickly in all three prognostic variables.
	The instability possesses a small scale structure with short wavelengths.
	
%--------------------------------------------------------------------------------------

	\section{Inviscid pseudo-incompressible waves}
	\label{sec:psinc}
	
	In this section we want to extend the model in order to incorporate anelastic effects as well
	which are omitted in the Boussinesq regime.
	The modulation equations for horizontally homogeneous waves 
	in the pseudo-incompressible regime exposed to an isothermal inviscid medium may be written as
	\begin{align}
		\label{eq:psinc_mod}	  
    	&\partial_Tk_z+\partial_Z\bigl(\hat{\omega}(k_z)+K_xu\bigr)=0,\notag\\
	  	&\partial_T\mathcal{A}+\partial_Z\bigl(\hat{\omega}'(k_z)\mathcal{A}\bigr)=0,\notag\\
    	\rho(Z)&\partial_Tu+\partial_Z\bigl(\hat{\omega}'(k_z)K_x\mathcal{A}\bigr)=0.
	\end{align}
	Here, in contrast to the Boussinesq model the background density 
	for the isothermal ambient atmosphere is a function of height,
	\begin{align}
		\label{eq:def_eta}
		\rho(Z)=\rho_0e^{\eta Z}\quad\text{with}\quad\eta=-\frac{7}{2}N^2.
	\end{align}
	It can be derived from the hydrostatic equation and the equation of state for ideal gases.
	The system \eqref{eq:psinc_mod} was introduced by \citet{Grimshaw1972}
	to describe nonlinear internal gravity waves 
	in an inviscid compressible fluid in the environment of
	a slowly varying medium. \citet{Achatz2010} demonstrated 
	that these are WKB solutions to the compressible Euler equations 
	and likewise to the pseudo-incompressible equations of \citet{Durran1989} 
	when applying the same asymptotic scaling.
	In contrast to the Boussinesq equations they can capture the well known anelastic 
	amplification of wave packets when they travel upwards.
	
	A derivation of these modulation equations is also presented in \citet{Schlutow2017b}.
	The Euler equations are scaled in terms of a small parameter 
	\begin{align}
		\varepsilon=L_r/H_\theta
	\end{align}
	where $L_r$ denotes a reference wave length and $H_\theta$ the potential temperature scale height.
	A distinguished limit for gravity waves is found to be
	\begin{align}
		\label{eq:psinc_disting}
		\mathit{Ma}=\mathit{O}(\varepsilon)\quad\text{and}\quad\mathit{Fr}=\mathit{O}(\varepsilon^{1/2}),
	\end{align}
	i.e. assuming a small Mach and Froude number.
	The nonlinear wave ansatz allows that the envelope, the phase 
	and the induced mean flow vary on the same length scale as the background given by $H_\theta$
	while the wave itself oscillates on a much shorter scale $L_r$.
	We call the ansatz nonlinear because the mean flow interacts 
	with the wave field already in the leading order.
	
	\subsection{Traveling wave solutions}
	To get rid of the varying coefficient in \eqref{eq:psinc_mod} 
	due to the height-depending background density we can again introduce 
	the specific wave action density via $\mathcal{A}(Z,T)=\rho(Z)a(Z,T)$. 
	When we now write our system in vector form,
	\begin{align}
		\label{eq:psinc_mod_vec}	  		
		\partial_Ty+\partial_ZF(y)=G(y),
	\end{align}
	an additional inhomogeneity $G$ appears on the right hand side because 
	of the derivative of the background density with respect to $Z$.
	The inhomogeneity is given by
	\begin{align}
		G(y)=\begin{pmatrix}
			0\\
			-\eta\hat{\omega}'(k_z)a\\
			-\eta\hat{\omega}'(k_z)K_xa
		\end{pmatrix}
	\end{align}
	and
	\begin{align}
		F(y)=\begin{pmatrix}
			\hat{\omega}(k_z)+K_xu\\
			\hat{\omega}'(k_z)a\\
			\hat{\omega}'(k_z)K_xa
		\end{pmatrix}
	\end{align}
	is the same flux as in \eqref{eq:boussi_mod_vec}.
	By the substitution with the specific wave action density 
	both the flux and the inhomogeneity become autonomous.
	Therefore, the modulation equation can only assume 
	a stationary solution $Y=(K_z,\,A,\,U)^\mathrm{T}$ if the inhomogeneity becomes zero.
	This is the case when $A=0$ or $\hat{\omega}'(K_z)=0$.	
	None of the according solutions corresponds to a gravity wave, so we omit them. 
	Alternatively, one can ask for traveling wave solutions, i.e. solutions of the form 
	$Y(Z,T)=Y(Z-CT)$ where $C$ is a constant that we will refer to as envelope velocity.

	A traveling wave solution $Y(\zeta)$, where $\zeta=Z-CT$, must then fulfill
	\begin{align}
		\label{eq:psinc_trawa}	  		
		\partial_\zeta\bigl(F(Y)-CY\bigr)=G(Y).
	\end{align}
	This system of ODEs can be be simplified to a single first-order ODE
	by integrating the first equation giving 
	\begin{align}
		\label{eq:psinc_u}
		U(K_z)&=-K_x^{-1}(\hat{\omega}(K_z)-CK_z)+U_0.
	\end{align}
	Integration of the third minus the second equation results in 
	\begin{align}
		\label{eq:psinc_a}	
		A(K_z)&=K_x^{-1}U(K_z)+A_0.
	\end{align}
	Equations \eqref{eq:psinc_u} and \eqref{eq:psinc_a} are diagnostic equations 
	for the specific wave action density and the mean-flow horizontal wind
	that we can treat as functions of $K_z$ now.
	The two integration constants $A_0$ and $U_0$ must be determined by some initial conditions.
	The single ODE reads then
	\begin{align}
		\label{eq:psinc_ode_kz}
		\partial_\zeta K_z=f(K_z)
	\end{align}
	where
	\begin{align}
		f(K_z)=\frac{2\eta}{K_x^2}\frac{\hat{\omega}'(K_z)A(K_z)}{{A^2}''(K_z)}
	\end{align}
	where primes denote the derivative with respect to $K_z$.
	\citet{Schlutow2017b} showed that \eqref{eq:psinc_ode_kz} 
	has solutions that represent upward traveling wave fronts.
	These solutions converge at the infinities to constant values which are
	called asymptotic rest states, so
	\begin{align}
		\lim_{\zeta\rightarrow\pm\infty}K_z(\zeta)=K_z^\pm.
	\end{align}
	The asymptotic rest states correspond to the zeros of $f$ in \eqref{eq:psinc_ode_kz}, 
	i.e. $\hat{\omega}'(K_z^-)=0$ and $A(K_z^+)=0$. 
	According to \eqref{eq:dispersion} and \eqref{eq:psinc_a}, 
	one finds that $K_z^-=0$ and $K_z^+<0$, respectively.
	
	The condition $\hat{\omega}'(K_z)<C$ for all $K_z\in(K_z^+,K_z^-)$
	is a necessity to guarantee that the denominator of $f$ does not assume a zero 
	such that $f$ stays bounded.
	Note in particular that $C>0$ so the waves travel upwards.
	Given a $K_x$ and $K_z^+$ \citet{Schlutow2017b} computed a critical envelope velocity 
	\begin{align}
		C_{crit}=\begin{cases}
			\hat{\omega}'(K_z^+/\sqrt{2})&\text{if }K_z^+<-K_x/\sqrt{2},\\
			\hat{\omega}'(K_z^+)&\text{if }K_z^+\geq-K_x/\sqrt{2}
		\end{cases}
	\end{align}
	which guarantees bounded solutions if one chooses $C>C_{crit}$.
	
	The complete solution $Y=(K_z,\,A,\,U)^\mathrm{T}$ to \eqref{eq:psinc_mod_vec} exhibits the asymptotic rest states 
	\begin{align}
		\label{eq:psinc_asy_rest}
		\lim_{\zeta\rightarrow\pm\infty}Y(\zeta)=Y^\pm=\bigl(K_z^\pm,\,A(K_z^\pm),\,U(K_z^\pm)\bigr)^\mathrm{T}
	\end{align}
	being computed in terms of \eqref{eq:psinc_a} and \eqref{eq:psinc_u}.
	In optics or signal processing one would identify this traveling wave front as a ``down-chirp''
	as it drops in frequency when observed at a fixed height.
	The mean-flow horizontal wind $u$ is accelerated in positive direction
	in a fashion such that a persistent mean flow is induced.
	Remarkably, traveling wave backs as solutions to \eqref{eq:psinc_mod} cannot exist 
	which we will show in appendix B.
	
	Also note that linear group and envelope velocity for nonlinear waves are not necessarily the same.
	We defined the linear group velocity as the derivative of the frequency with respect to the wave number.
	It actually only coincides with the envelope velocity if the wave packet is quasi-monochromatic.
	Quasi-monochromatic waves may be found in linear and weakly nonlinear theory.
	Here in contrast we allow the vertical wavenumber to vary on the same order as the amplitude and the mean flow,
	so the prerequisite for monochronicity is generally not given.
	Instead $\hat{\omega}'$ describes the velocity of the group only locally in the vicinity of a given $k_z$.
	For a detailed derivation and discussion we refer to \citet[pp 86]{Majda2003}.

	\subsection{Stability of the inviscid traveling wave front}
	\label{sec:psinc_stab}
	
	To assess the stability of the inviscid pseudo-incompressible traveling wave fronts
	we first recast \eqref{eq:psinc_mod_vec} 
	in translational coordinates
	\begin{align}
		\label{eq:psinc_trans_coord}
		\zeta=Z-CT\quad\text{and}\quad\tau=T.
	\end{align}
	We benefit from this coordinate transformation 
	because the traveling wave front ${Y=Y(\zeta)}$ is a stationary solution in this system.
	Second, we linearize the transformed equation \eqref{eq:psinc_mod_vec}
	around the traveling wave solution $Y=(K_z,\,A,\,U)^\mathrm{T}$, giving
	\begin{align}
		\label{eq:psinc_mod_lin}
		\partial_\tau y+\partial_\zeta\bigl(\mathrm{D}F(Y)y-Cy\bigr)=\mathrm{D}G(Y)y
	\end{align}
	with the Jacobians
	\begin{align}
	  \mathrm{D}F(Y)=\begin{pmatrix}
	    \hat\omega'(K_z)& 0& K_x\\
	    A\hat{\omega}''(K_z)& \hat\omega'(K_z)& 0\\
	    K_xA\hat{\omega}''(K_z)& K_x\hat\omega'(K_z)& 0
	  \end{pmatrix}
	\end{align}
	and
	\begin{align}
	  \mathrm{D}G(Y)=\begin{pmatrix}
	    0& 0& 0\\
	    -\eta A\hat{\omega}''(K_z)& -\eta\hat{\omega}'(K_z)& 0\\
	    -\eta K_xA\hat{\omega}''(K_z)& -\eta K_x\hat{\omega}'(K_z)& 0
	  \end{pmatrix}.
	\end{align}
	Since the Jacobians of the flux and the inhomogeneity are independent 
	of the time variable $\tau$ by autonomy and the stationarity of $Y$,
	equation \eqref{eq:psinc_mod_lin} must have solutions of the form
	\begin{align}
		y(\zeta,\tau)=y(\zeta)e^{\lambda\tau}
	\end{align}
	which yields an eigenvalue problem $(\mathcal{L}_Y-\lambda)y=0$.
	The linear operator is given by
	\begin{align}
		\label{eq:psinc_lin_op}
		\mathcal{L}_Y=\mathrm{D}G(Y)-\partial_\zeta\mathrm{D}F(Y)-\bigl(\mathrm{D}F(Y)-C\bigr)\partial_\zeta.
	\end{align}

	As for the Boussinesq plane waves we also translated the problem of linear stability 
	into the task of finding the spectrum of a linear operator.
	However, in contrast to the plane waves this operator exhibits varying coefficients 
	due to the spatial dependency of the traveling wave front $Y=Y(\zeta)$.
	In conclusion, we cannot compute the spectrum straightforwardly 
	by a Fourier transformation as we did for the plane waves.
	
	But fortunately, analytical progress is possible utilizing the asymptotic rest states.
	The idea is essentially to approximate the operator using these states by an asymptotic operator
	for which the spectrum is easy to compute. 
	We can connect the asymptotic to the original operator by Fredholm theory. 
	For a detailed introduction on this topic we refer to \citet{Sandstede2002,Kapitula2013}.
	
	Let $\mathcal{L}$ be a closed linear operator defined 
	on some Hilbert space $X$ with a domain being dense in $X$.	
	The operator $\mathcal{L}$ is called Fredholm if the dimension of its kernel 
	and the codimension of its range are finite. 
	The Fredholm index is then defined by
	\begin{align}
		\label{eq:psinc_def_fred}
		\operatorname{ind}\mathcal{L}=\dim(\ker\mathcal{L})
		-\operatorname{codim}(\operatorname{range}\mathcal{L}).
	\end{align}
	In terms of the above definition we can separate the spectrum of $\mathcal{L}$
	into two sets, the essential and the point spectrum.
	\begin{itemize}
		\item The essential spectrum of $\mathcal{L}$ is the set 
		of all $\lambda\in\mathbb{C}$ such that one of the following is true:
		either $\mathcal{L}-\lambda$ is not Fredholm, or $\mathcal{L}-\lambda$ 
		is Fredholm but $\operatorname{ind}(\mathcal{L}-\lambda)\neq 0$.
		\item The point spectrum of $\mathcal{L}$ is the set of all $\lambda\in\mathbb{C}$ 
		such that $\mathcal{L}-\lambda$ is Fredholm and $\operatorname{ind}(\mathcal{L}-\lambda)=0$ 
		but $\mathcal{L}-\lambda$ is not invertible.
	\end{itemize}
	
	With the aid of these definitions one can 
	now apply Weyl's essential spectrum theorem \citep[pp 29]{Kapitula2013}
	which states that a well-behaved perturbation $\mathcal{L}$ 
	of a Fredholm operator $\mathcal{L}_0$ is Fredholm, 
	that the Fredholm indices are the same, and that $\mathcal{L}$ 
	and $\mathcal{L}_0$ share the same essential spectrum.

	Let us define the asymptotic operator to \eqref{eq:psinc_lin_op} as
	\begin{align}
		\mathcal{L}_\infty=\begin{cases}
			\mathcal{L}_{Y^-}&\text{if }\zeta<0\\
			\mathcal{L}_{Y^+}&\text{if }\zeta\geq 0.
		\end{cases}
	\end{align} 
	One can readily show that the original operator $\mathcal{L}_Y$
	is indeed a well-behaved perturbation of $\mathcal{L}_\infty$ 
	in the required sense \citep[pp 39]{Kapitula2013}.	
	Therefore, we can compute the essential spectrum of $\mathcal{L}_\infty$ 
	and make use of Weyl's essential spectrum theorem by 
	identifying it with the essential spectrum of $\mathcal{L}_Y$.
	
	The asymptotic eigenvalue equation $(\mathcal{L}_\infty-\lambda)y=0$ can be recast as the ODE
	\begin{align}
		\partial_\zeta y=B(\lambda;\zeta)y.
	\end{align}
	having a piecewise constant coefficient matrix generated by 
	\begin{align}
		B(\lambda;\zeta)=\begin{cases}
			B^-(\lambda)&\text{if }\zeta<0\\
			B^+(\lambda)&\text{if }\zeta\geq 0.
		\end{cases}
	\end{align}
	From \eqref{eq:psinc_lin_op} we find that
	\begin{align}
		\label{eq:psinc_bpm}
		B^\pm(\lambda)=(\mathrm{D}F(Y^\pm)-C)^{-1}(\mathrm{D}G(Y^\pm)-\lambda).
	\end{align}
	The eigenvalues $\nu^\pm(\lambda)$ of $B^\pm(\lambda)$ 
	contain all the information we need to characterize the essential spectrum.
	The operators $\mathcal{L}_\infty-\lambda$ 
	and hence $\mathcal{L}_Y-\lambda$ are Fredholm if, and only if,
	all eigenvalues $\nu^\pm(\lambda)$ have non-zero real part, 
	i.e. $B^\pm(\lambda)$ is hyperbolic \citep{Sandstede2002}.
	The Fredholm index can be written as
	\begin{align}
		\label{eq:psinc_ind}
		\operatorname{ind}(\mathcal{L}_Y-\lambda)=i_-(\lambda)-i_+(\lambda)
	\end{align}
	where $i_\pm$ are the Morse indices of $B^\pm(\lambda)$.
	The Morse index of a hyperbolic matrix is the dimension of its unstable subspace.
	So, if for a $\lambda\in\mathbb{C}$ the operator is Fredholm and the Morse indices differ, 
	i.e. $i_-(\lambda)\neq i_+(\lambda)$, 
	then $\lambda$ is in the essential spectrum of $\mathcal{L}_Y$.
		
	We can circumnavigate the computation of the Morse indices for every $\lambda\in\mathbb{C}$
	by investigating the boundaries of the essential spectrum directly.
	The essential spectrum is delimited by curves where the $\lambda$'s 
	are such that the matrices $B^\pm(\lambda)$ fail to be hyperbolic,
	that is where some eigenvalues become purely imaginary, 
	so $\nu^\pm(\lambda)=i\mu$ and $\mu\in\mathbb{R}$.
	These $\lambda$'s satisfy the dispersion relation
	\begin{align}
		\label{eq:psinc_det}
		\det(B^\pm(\lambda)-i\mu)=0.
	\end{align}
	Plugging \eqref{eq:psinc_bpm} into \eqref{eq:psinc_det} and simplifying provides
	\begin{align}
		\det\bigl(\mathrm{D}G(Y^\pm)-\lambda-i\mu(\mathrm{D}F(Y^\pm)-C)\bigr)=0.
	\end{align}
	We substitute the asymptotic rest states from \eqref{eq:psinc_asy_rest} 	
	and solve for $\lambda$ 
	which results in six curves in the complex plane being parametrized by $\mu$
	\begin{align}
		\label{eq:la1p}
		\lambda_1^+(\mu)&=iC\mu\\
		\label{eq:la2p}
		\lambda_2^+(\mu)&=-i(\hat{\omega}'(K_z^+)-C)\mu\\
		\label{eq:la3p}
		\lambda_3^+(\mu)&=-\eta\hat{\omega}'(K_z^+)-i(\hat{\omega}'(K_z^+)-C)\mu\\
		\label{eq:la1m}
		\lambda_1^-(\mu)&=iC\mu\\
		\label{eq:la2m}
		\lambda_2^-(\mu)&=+\sqrt{NA(K_z^-)(\mu^2-i\eta\mu)}+iC\mu\\
		\label{eq:la3m}
		\lambda_3^-(\mu)&=-\sqrt{NA(K_z^-)(\mu^2-i\eta\mu)}+iC\mu
	\end{align}
	where we also identified and replaced $\hat{\omega}''(K_z^-=0)=-N/K_x^2$.
	These curves are denoted as Fredholm borders. 
	They are the boundaries of the regions in the complex plane 
	where the operator $\mathcal{L}_Y-\lambda$ is Fredholm.
	\begin{figure}[ht]
		\begin{center}
			\input{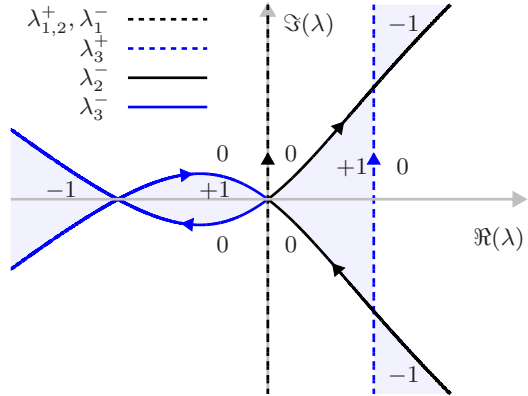}
		\end{center}
		\caption{Essential spectrum of $\mathcal{L}_Y$ with Fredholm indices 
		and borders for the inviscid pseudo-incompressible traveling wave front.}
		\label{fig:psinc_spec}
	\end{figure}
	For a typical set of parameters defining a pseudo-incompressible inviscid traveling wave front, 
	the Fredholm borders are plotted in \mbox{figure \ref{fig:psinc_spec}}. 
	Arrows depict their orientations which are determined by the direction of increasing $\mu$.
	As $\lambda$ crosses a Fredholm border the Fredholm index may change.
	And indeed, one can prove that $\operatorname{ind}(\mathcal{L}_Y-\lambda)$ 
	\begin{itemize}
		\item increases by 1 when crossing the graph of $\lambda^+(\mu)$ 
		from right to left with regard to its orientation.
		\item decreases by 1 when crossing the graph of $\lambda^-(\mu)$ 
		from right to left with regard to its orientation.
	\end{itemize}

	An asymptotic analysis reveals that real valued $\lambda\gg 1$ are not in the essential spectrum.
	This means that the Fredholm index in the regions containing these $\lambda$'s is zero.
	Since we know the Fredholm index in one region, 
	we can compute the other indices of the remaining regions by the rule from above.
	The regions with Fredholm index other than zero belong to the essential spectrum.
	
	Now, as we know the essential spectrum of the operator, we can analyze it with respect to 
	instabilities which means to check for $\Re(\lambda)>0$ in the essential spectrum.
	\begin{figure}[ht]
		\centering
		\input{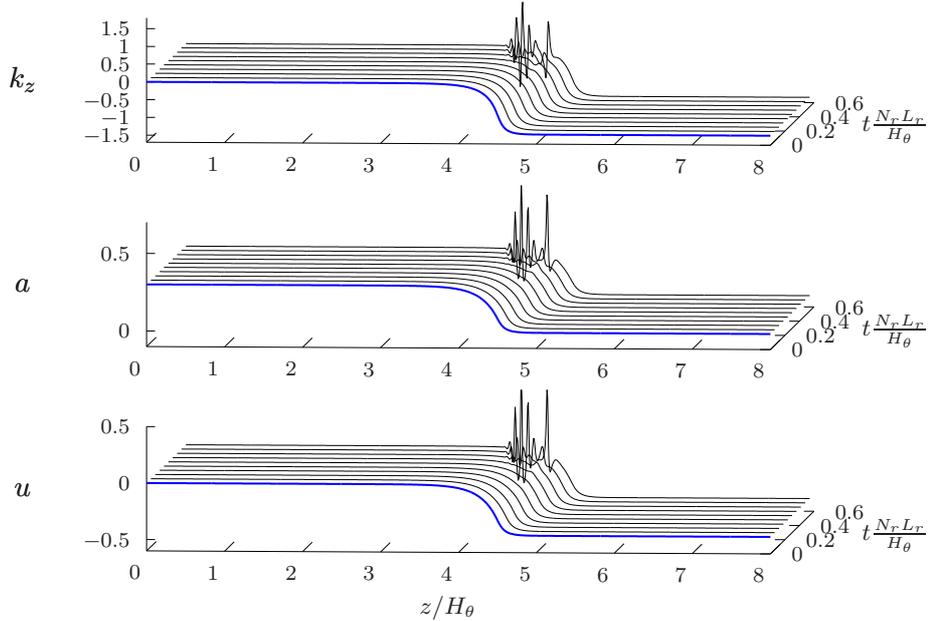}
		\caption{Numerical simulation of the upward-traveling 
		pseudo-incompressible inviscid  wave showing absolute instability.
		It blows up in under 15\,min.}
		\label{fig:psinc_adistrag}
	\end{figure}
%	%
	We can see in figure \ref{fig:psinc_spec} that, for this particular choice of parameters, 
	the spectrum reaches far into the right half of the complex plane. 
	And indeed we find from \eqref{eq:la3p} that $\lambda_3^+$
	always has positive real part $-\eta\hat{\omega}'(K_z^+)$ 
	because $\eta<0$ as the background
	density decreases with height and $\hat{\omega}'(K_z^+)>0$ 
	which is a prerequisite for the existence of the traveling wave front.
	Also, it can be shown that $\Re(\lambda_2^-)>0$ independent of the choice of parameters. 
	We can therefore deduce that the traveling wave front is spectrally unstable 
	with respect to the modulation equations 
	for every set of parameters defining the wave and the background.

	\subsection{Numerical simulation of the inviscid traveling wave front}
	\label{sec:psinc_num}
	
	We want to compute the evolution of the essential spectrum instability of 
	the  inviscid traveling wave front numerically to learn about its dynamics.
	For this purpose, the modulation equations \eqref{eq:psinc_mod} are integrated in time 
	with the numerical method presented in appendix A.
	The parameters are set to $N=1.7$, $K_x=1.5$ and $C=0.8$.
	By integrating equation \eqref{eq:psinc_ode_kz} numerically we obtain the initial conditions.
	
	The numerical results of the time integration are plotted in figure \ref{fig:psinc_adistrag}.
	We labeled the axes with the dimensional height $z$ and time $t$ where
	$N_r^2=g/H_\theta$ and $g$ the Earth's acceleration according to \citet{Schlutow2017b}.
	An unstable wave packet grows rapidly directly behind the front.
	The simulation is terminated by the numerics being unable to handle the small-scale instability.
	For typical values for the mesosphere, $H_\theta=20$\,km and $L_r=1$\,km,
	we can estimate that after only 15\,min the wave blows up.

%--------------------------------------------------------------------------------------
	
	\section{Dissipative Grimshaw waves}
	\label{sec:grimshaw}

	So far, we have seen in section \ref{sec:bouss} and \ref{sec:psinc} that both
	the inviscid Boussinesq plane wave and the inviscid pseudo-incompressible traveling wave front 
	exhibit instabilities which grow rapidly in our numerical simulations.
	They also suffer from an unphysical spectrum.
	Both spectra allow for unstable modes with arbitrarily large instability growth rates
	which makes instabilities potentially catastrophic.
	According to \citet[p 61]{Kapitula2013} the operators are not well-posed.
		
	This problem originates intrinsically from the nature of the Euler equations that we approximated.
	By omitting diffusive effects we removed the damping 
	which is especially efficient in suppressing high-frequency perturbations.
	
	In the Boussinesq case, \citet{Fetecau2011} proposed a regularization 
	by including higher order terms to the dispersion relation.
	In contrast to this \textit{ad hoc} method we want to introduce
	a regularization to the pseudo-incompressible regime 
	by incorporating dissipative, instead of dispersive, effects in an asymptotic fashion.

	The modulation equations for nonlinear gravity waves 
	including dissipation according to \citet{Grimshaw1974} 
	assuming horizontal homogeneity and isothermal background are given by
	\begin{align}
		\label{eq:grimshaw_mod}
	  	\partial_T&k_z+\partial_Z\bigl(\hat{\omega}(k_z)+K_xu\bigr)=0\\
	  	\partial_T&\mathcal{A}+\partial_Z\bigl(\hat{\omega}'(k_z)\mathcal{A}\bigr)=-\Lambda k^2(k_z)\mathcal{A}\\
	  	\rho(Z)\partial_T&u+\partial_Z\bigl(K_x\hat{\omega}'(k_z)\mathcal{A}\bigr)=0.
	\end{align}
	Note that the only difference to the inviscid case of the previous section 
	is an additional term on the right hand side that accounts for the dissipative effects.
	It consists of the squared norm $k^2(k_z)=K_x^2+k_z^2$ as function of $k_z$, the wave action density
	and a variable $\Lambda$ that comprises the effects of viscosity and thermal conductivity. 
	
	The dissipative modulation equations can be derived 
	from the compressible Navier-Stokes equations in the same fashion 
	as if one starts from the compressible Euler equations. 
	The small parameter is likewise the ratio of reference wave length 
	and potential temperature scale height. 
	The enhanced distinguished limit (cf. equation \ref{eq:psinc_disting}) 
	from the asymptotic scaling reads then
	\begin{align}
		\mathit{Ma}=\mathit{O}(\varepsilon),~
		\mathit{Fr}=\mathit{O}(\varepsilon^{1/2}),~
		\mathit{Re}=\mathit{O}(\varepsilon^{-1})~
		\text{and}~\mathit{Pr}=\mathit{O}(1).
	\end{align}
	So, we assume a large Reynolds number 
	and by the Prandtl number that the viscosity is of the same order as the thermal conductivity.
	This scaling is realistic for two regions. Either the lower atmosphere 
	when we treat the equations as Reynolds averaged with a simple turbulence model 
	such that the Reynolds number represents eddy viscosity.
	Or for the regions of the higher atmosphere, i.e. the mesosphere and lower thermosphere, 
	where molecular viscosity becomes large.
	
	The nonlinear WKB wave ansatz is the very same as for the inviscid case.
	The phase, wave envelope, mean flow and background vary on the same length scale $H_\theta$.
	Mean flow and wave field interact in leading order.
	A thorough derivation of the resulting modulation equations 
	can be found in \citet{Grimshaw1974}
	which is indeed a consistent dissipative extension 
	of \citet{Schlutow2017b} with regard to the scaling presented here.
	
	The variable $\Lambda$ is a function of altitude $Z$ 
	and is found to be proportional to the inverse background density
	which implies that the impact of dissipative effects increases with height. 
	However, we treat $\Lambda$ as a constant henceforth
	in order to continue with our analytical approach.
	On the one hand, we are interested in a regularization of the inviscid traveling wave fronts.
	Setting $\Lambda$ constant enables us to construct an autonomous flux and inhomogeneity
	giving rise to traveling wave solutions with asymptotic rest states similar to the inviscid waves.
	It is reasonable to assume that the non-autonomy caused by a varying $\Lambda$
	may change the solution but the stability analysis based on the asymptotic rest states
	is most likely untouched though.
	
	On the other hand, \citet{Grimshaw1974} argues that this assumption is not a great restriction
	as a varying dissipation is only of importance in the vicinity of critical levels 
	where the extrinsic frequency approaches zero such that the wave action density
	amplifies. In our scenario there are no critical levels present.
	\citet{Grimshaw1974} also tested the influence of varying versus constant $\Lambda$ numerically 
	and found no significant difference.
	
	\subsection{The traveling wave solution}
	
	To construct traveling wave solutions to the dissipative nonlinear modulation equations \eqref{eq:grimshaw_mod}
	we exploit the transformation generated by the specific wave action density $a(Z,T)=\rho^{-1}(Z)\mathcal{A}(Z,T)$
	where $\rho$ is the same exponentially decreasing background density as before.
	This transformation removes the $Z$-dependencies, 
	such that the coefficients of the resulting system become constant.
	Written in vector form the system reads
	\begin{align}
		\label{eq:grimshaw_mod_vec}
	  	\partial_Ty+\partial_ZF(y)=G(y)
	\end{align}
	for the solution vector $y=(k_z,\,a,\,u)^\mathrm{T}$.
	The flux and the inhomogeneity are given by
	\begin{align}
		F(y)=\begin{pmatrix}
			\hat{\omega}(k_z)+K_xu\\
			\hat{\omega}'(k_z)a\\
			\hat{\omega}'(k_z)K_xa
		\end{pmatrix}
	\end{align}
	and
	\begin{align}
		G(y)=\begin{pmatrix}
			0\\
			-\eta\hat{\omega}'(k_z)a-\Lambda k^2(k_z)a\\
			-\eta\hat{\omega}'(k_z)K_xa
		\end{pmatrix},
	\end{align}
	respectively.
	A traveling wave solution $Y(\zeta)=(K_z,\,A,\,U)^\mathrm{T}$, where $\zeta=Z-CT$, fulfills 
	\begin{align}
	  \label{eq:grimshaw_ode}
	  \partial_\zeta \bigl(F(Y)-CY\bigr)=G(Y)
	\end{align}
	that states a three-dimensional ODE.
	We can integrate the first equation of the system, 
	since the right hand side is zero,
	which yields a diagnostic equation for the mean-flow horizontal wind
	\begin{align}
		\label{eq:grimshaw_u}
		U(K_z)=-K_x^{-1}\bigl(\hat{\omega}(K_z)-CK_z\bigr)+U_0
	\end{align}
	with an integration constant $U_0$ that depends on the initial conditions.
	Thus, we reduce the dimension to two and $U$ is now treated as a function of $K_z$. 
	With a mild abuse of notation we rewrite the solution vector to $Y=(K_z,\,A)^\mathrm{T}$
	and the system of ODEs \eqref{eq:grimshaw_ode} becomes
	\begin{align}
		\label{eq:grimshaw_ode2}
	  \partial_\zeta f(Y)=g(Y)
	\end{align}
	The flux and inhomogeneity are updated by substituting \eqref{eq:grimshaw_u}, so
	\begin{align}
	  f(Y)=\begin{pmatrix}
	    (\hat{\omega}'(K_z)-C)A\\
	    K_x\hat{\omega}'(K_z)A-CU(K_z)
	  \end{pmatrix} 
	\end{align}
	and
	\begin{align}
	  g(Y)=\begin{pmatrix}
	    -\Lambda k^2(K_z)A-\eta \hat{\omega}'(K_z)A\\
	    -\eta K_x\hat{\omega}'(K_z)A
	  \end{pmatrix}.
	\end{align}
	The ODE \eqref{eq:grimshaw_ode2} has equilibria if the inhomogeneity vanishes.
	Those points are ${Y=(K_z,\,0)}$, i.e. all points on the $K_z$-axis in the two-dimensional phase space.
	So, there may be solutions with two asymptotic rest states, $Y^+$ and $Y^-$, on the $K_z$-axis
	where the specific wave action density vanishes at the infinities
	and the vertical wave number approaches some constant values.
	The solution is then the trajectory in phase space 
	parametrized by $\zeta$ that connects those two equilibria.

	From a point of view of dynamical systems we can investigate the character of the equilibria 
	by linearizing the ODE \eqref{eq:grimshaw_ode2} around these states,
	which yields
	\begin{align}
		\label{eq:grimshaw_char_ode}
	  \mathrm{D}f(Y^\pm)\partial_\zeta Y = \mathrm{D}g(Y^\pm)Y,
	\end{align}
	in order to find out if an equilibrium is an attractor or a repeller.
	The Jacobians in \eqref{eq:grimshaw_char_ode} can be computed as
	\begin{align}
	  \mathrm{D}f(Y)&=\begin{pmatrix}
	    A\hat{\omega}''(K_z)& \hat{\omega}'(K_z)-C\\
	    K_xA\hat{\omega}''(K_z)-CU'(k_z)& K_x\hat{\omega}'(K_z)
	  \end{pmatrix}
	 \end{align}
	 and
	 \begin{align}
	  \mathrm{D}g(Y)&=\begin{pmatrix}
	    -2\Lambda K_zA-\eta\hat{\omega}''(K_z)A& -\Lambda k^2(K_z)-\eta\hat{\omega}'(K_z)\\
	    -\eta K_xA\hat{\omega}''(K_z)& -\eta K_x\hat{\omega}'(K_z)
	  \end{pmatrix}.
	\end{align}	
	Evaluating them at the equilibria yields
	\begin{align}
		\mathrm{D}f(Y^\pm)&=\begin{pmatrix}
		  	0& \hat{\omega}'(K_z^\pm)-C\\
		  	-CU'(K_z^\pm)& K_x\hat{\omega}'(K_z^\pm)
		\end{pmatrix}
	\end{align}
	 and
	 \begin{align}
		\mathrm{D}g(Y^\pm)&=\begin{pmatrix}
		  	0& -\Lambda k^2(K_z^\pm)-\eta\hat{\omega}'(K_z^\pm)\\
		  	0& -\eta K_x\hat{\omega}'(K_z^\pm)
		\end{pmatrix}.
	\end{align}
	Equation \eqref{eq:grimshaw_char_ode} has solutions of the form $Y(\zeta)=Ye^{\sigma\zeta}$
	giving us the characteristic equation $\sigma^2-p\sigma=0$ with the coefficient being also a root
	\begin{align}
		\label{eq:grimshaw_p}
	  	p(K_z)=-\frac{\Lambda k^2(K_z)+\eta\hat{\omega}'(K_z)}{\Delta(K_z)}.
	\end{align}
	Here, we introduced $\Delta(K_z)=\hat{\omega}'(K_z)-C$ 
	as the difference of linear group and the envelope velocity.
	Equilibria are attractive if $p<0$ and repellent if $p>0$.
	In particular we can identify that $p(K_z^-)>0$ and $p(K_z^+)<0$ must be true
	if we want to associate $K_z^\pm$ with the asymptotic rest states at $\zeta=\pm\infty$.
	The sign of $p$ depends partly on the denominator $\Delta$. 
	
	For downward-traveling wave packets, i.e. $C<0$, the sign of $\Delta$ is positive 
	as $\hat{\omega}'(K_z^\pm)>0$ for $K_z^\pm<0$ which we anticipate here and explain subsequently.
	For upward-traveling wave packets, i.e. $C>0$, we must make sure 
	that $\Delta$ does not assume a zero.
	Since $\hat{\omega}'$ has a global maximum at $-K_x/\sqrt{2}$, 
	it proved useful to choose
	\begin{align}
		C>\hat{\omega}'(-K_x/\sqrt{2})=\frac{2}{3\sqrt{3}}\frac{N}{K_x}.
	\end{align}	 
	By this choice $\Delta$ is strictly negative.
	
	It can be also established by continuity that there must be a $\bar K_z$
	between $K_z^-$ and $K_z^+$ such that $p(\bar K_z)=0$.
	When the equilibrium passes this point,
	$p$ switches the sign being determined by the numerator of \eqref{eq:grimshaw_p}
	as the denominator, $\Delta$, does not assume a zero.

	To evaluate the trajectory numerically we use the explicit fourth-order Runge-Kutta method
	and rearrange \eqref{eq:grimshaw_ode2} by applying the chain rule to
	\begin{align}
		\label{eq:grimshaw_ode3}
	  	\partial_\zeta Y =h(Y)\quad\text{and}\quad h(Y)=\mathrm{D}f^{-1}(Y)g(Y).
	\end{align}
	By matrix inversion we obtain
	\begin{align}
	  \mathrm{D}f^{-1}(Y)=
	  \frac{1}{\det\bigl(\mathrm{D}f(Y)\bigr)}
	  \begin{pmatrix}
	    K_x\hat{\omega}'(K_z)& -\Delta(K_z)\\
	    -K_xA\hat{\omega}''(K_z)-CU'(K_z)& A\hat{\omega}''(K_z)
	  \end{pmatrix}
	\end{align}
	where 
	\begin{align}
	  \det\bigl(\mathrm{D}f(Y)\bigr)=-CK_x^{-1}\left(K_x^2A\hat{\omega}''(K_z)-\Delta^2(K_z)\right).
	\end{align}
	In particular $\det\bigl(\mathrm{D}f(Y)\bigr)=0$ gives a singularity if
	\begin{align}
		\label{eq:dissi_sin}
	  	A_\mathrm{sin}(K_z)=\frac{\Delta^2(K_z)}{K_x^2\hat{\omega}''(K_z)}
	\end{align}
	which parametrizes a curve in the phase space.
	Horizontal isoclines, which are curves in phase space where the vectors 
	are horizontal or formally $h=(h_{K_z},0)^\mathrm{T}$, are derived from \eqref{eq:grimshaw_ode3} being
	\begin{align}
		\label{eq:dissi_hori_iso}
	  A_\mathrm{iso}(K_z)=CU'(K_z)\frac{\Lambda k^2(K_z)+\eta\hat{\omega}'(K_z)}{K_x\hat{\omega}''(K_z)\Lambda k^2(K_z)}.
	\end{align}
	Three vertical isocline where $h=(0,h_A)^\mathrm{T}$ turn out 
	to be vertical straight lines that cross the $K_z$-axis at
	\begin{align}
		\label{eq:dissi_vert_iso}
	  	K_z^\mathrm{iso}&=0,\quad\pm\sqrt{-\eta\Lambda^{-1}C-K_x^2}.
	\end{align} 
	\begin{figure}[ht]
		\centering
	  	\input{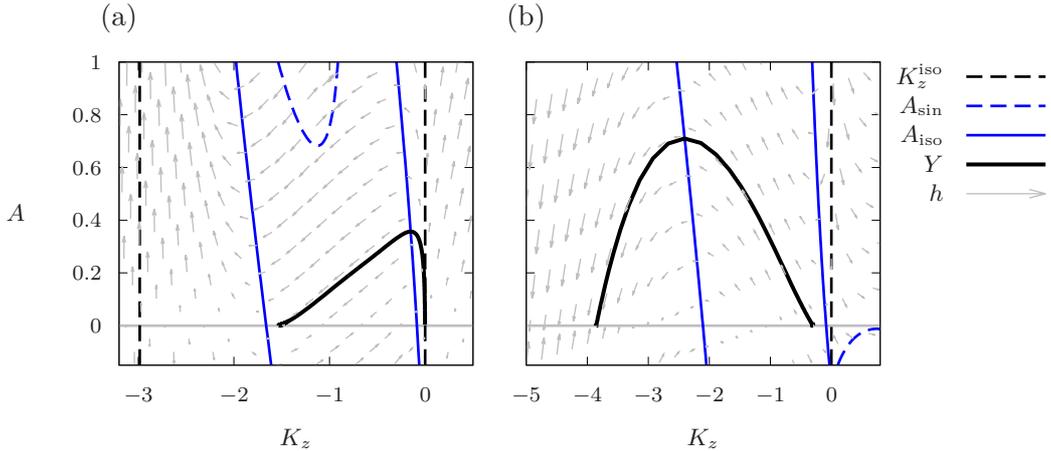}
	  	\caption{Phase portraits for dissipative Grimshaw traveling wave packets and solution trajectory.
	  	Panel (a): Upward-traveling wave where $C>0$. Panel (b): Downward-traveling wave for $C<0$.}
	  	\label{fig:grimshaw_phase_portrait}
	\end{figure}
%	%
	Prototypical phase portraits together with the trajectories computed numerically 
	by the Runge-Kutta method
	are depicted in figure \ref{fig:grimshaw_phase_portrait} for an upward and a downward-traveling wave packet, respectively.
	In terms of the previous considerations we can exploit phase plane analysis 
	to show that \eqref{eq:grimshaw_ode2} is indeed solved by trajectories of the form of heteroclinic orbits.
	In both plots the $K_z$-axis consists of equilibria (grey line) and the $A$-axis is an isocline denoted by $K_z^\mathrm{iso}$
	where the vectors are vertical.
	This implies that a trajectory initialized in the second quadrant, i.e. where $K_z<0$ and $A>0$, stays in it.
	By the symmetry of the problem and the choice of the branch of $\hat{\omega}$,
	it is sufficient, without loss of generality, to focus on the second quadrant only.
	
	Let us consider the upward-traveling wave ($C>0$) 
	which is shown in panel (a) of figure \ref{fig:grimshaw_phase_portrait}.
	Besides the aforementioned vertical isocline, we find a second one and 
	two additional branches of the horizontal isocline $A_\mathrm{iso}$. 
	These four curves delimit three regions outside of the singularity 
	where the sign of the components of $h$ are constant.
	In particular the region outside of the singularity 
	between the two branches of $A_\mathrm{iso}$ has vectors
	that possess only negative horizontal and negative vertical components.
	So, for every $K_z^+$ on the $K_z$-axis between the two branches of $A_\mathrm{iso}$
	we already know that by $p(K_z^+)<0$ we obtain an attractor
	which implies that there is a trajectory ending in the point $(K_z^+,0)$ for $\zeta\rightarrow+\infty$ 
	that we can trace backwards in $\zeta$.
	This point is connected by the trajectory with some point on the right branch of $A_\mathrm{iso}$
	if $K_z^+$ is far enough to the right such that it is clear from the singularity.
	All vectors, $h$, between the right branch of $A_\mathrm{iso}$ and the $A$-axis point upwards and to the left.
	Hence, there must be a $K_z^-$ in this region
	which is a repeller due to $p(K_z^-)>0$ such that it connects 
	the crossing point on the right branch of $A_\mathrm{iso}$ with $(K_z^-,0)$ when we keep tracing back. 
	The resulting concatenated trajectory is then a heteroclinic orbit that we refer to as wave packet
	because the specific wave action density is of finite extent in contrast to the traveling wave front
	in the inviscid case.
	
	Let us now consider the case $C<0$
	corresponding to downward-traveling wave packets.
	As depicted in figure \ref{fig:grimshaw_phase_portrait} on the right panel
	we also find the two branches of $A_\mathrm{iso}$.
	However, the character of the equilibrium points between these two changes.
	In contrast to the upward-traveling wave, these equilibria are repellent, so $p(K_z^-)>0$.
	The components of $h$ possess constant sign in the region between the two branches of  
	$A_\mathrm{iso}$ and therefore all vectors point upwards and to the left.
	A trajectory that starts at $K_z^-$ sufficiently close to the left branch of $A_\mathrm{iso}$ 
	must cross it at some point when integrating forwards.
	All vectors to its left point downwards and hence
	the trajectory must land on an equilibrium point $K_z^+$
	giving us a heteroclinic orbit in accordance with a downward-traveling wave packet.

	In conclusion, we found both up and downward-traveling wave packets solving the dissipative Grimshaw
	modulation equations. An illustration of these waves 
	will be given in section \ref{sec:grimshaw_num} 
	when we study the evolution of instabilities which is presented in the following section.

	\subsection{Stability of the dissipative Grimshaw traveling wave packets}
	\label{sec:grimshaw_stab}
	
	In this section we will investigate the stability of the dissipative traveling wave packets.
	We linearize the modulation equations \eqref{eq:grimshaw_mod_vec} around the traveling wave solutions 
	in the translational coordinate system presented in \eqref{eq:psinc_trans_coord}
	\begin{align}
		\label{eq:grimshaw_lin}
		\partial_\tau y+\partial_\zeta\bigl(\mathrm{D}F(Y)y-Cy\bigr)=\mathrm{D}G(Y)y
	\end{align}
	with the Jacobian matrices
	\begin{align}
	  \mathrm{D}F(Y)=\begin{pmatrix}
	    \hat\omega'(K_z)& 0& K_x\\
	    A\hat{\omega}''(K_z)& \hat\omega'(K_z)& 0\\
	    K_xA\hat{\omega}''(K_z)& K_x\hat\omega'(K_z)& 0
	  \end{pmatrix}
	\end{align}
	and
	\begin{align}
	  \mathrm{D}G(Y)=\begin{pmatrix}
	    0& 0& 0\\
	    -2\Lambda K_zA-\eta A\hat{\omega}''(K_z)& -\Lambda k^2(K_z)-\eta\hat{\omega}'(K_z)& 0\\
	    -\eta K_xA\hat{\omega}''(K_z)& -\eta K_x\hat{\omega}'(K_z)& 0
	  \end{pmatrix}.
	\end{align}

	Since the traveling wave solution $Y(\zeta)$ is stationary in the translational coordinates 
	and the system is autonomous, equation \eqref{eq:grimshaw_lin} 
	must have solutions of the form
	\begin{align}
	  y(\zeta,\tau)=y(\zeta)e^{\lambda\tau}
	\end{align}
	which yields an eigenvalue problem $(\mathcal{L}_Y-\lambda)y=0$.
	The linear operator is given by
	\begin{align}
		\label{eq:grimshaw_lin_op}
		\mathcal{L}_Y=\mathrm{D}G(Y)-\partial_\zeta\mathrm{D}F(Y)-\bigl(\mathrm{D}F(Y)-C\bigr)\partial_\zeta\,.
	\end{align}
	We proceed with the same method being presented in section \ref{sec:psinc_stab}.
	As the traveling wave solution converges to its asymptotic rest states at the infinities,
	we can construct an asymptotic operator
	\begin{align}
		\mathcal{L}_\infty=\begin{cases}
			\mathcal{L}_{Y^-}&\text{if }\zeta<0\\
			\mathcal{L}_{Y^+}&\text{if }\zeta\geq 0
		\end{cases}
	\end{align}
	that exhibits the same essential spectrum as $\mathcal{L}_Y$ itself.

	The new eigenvalue problem $(\mathcal{L}_\infty-\lambda)y=0$ is rearranged to
	a first-order ODE, 
	\begin{align}
		\partial_\zeta y=B(\lambda;\zeta)y\,,
	\end{align}		
	with the piecewise constant coefficient matrix
	\begin{align}
		B(\lambda;\zeta)=\begin{cases}
			B^-(\lambda)&\text{if }\zeta<0\\
			B^+(\lambda)&\text{if }\zeta\geq 0
		\end{cases}
	\end{align}
	and 
	\begin{align}
		\label{eq:grimshaw_bpm}
		B^\pm(\lambda)=(\mathrm{D}F(Y^\pm)-C)^{-1}(\mathrm{D}G(Y^\pm)-\lambda).
	\end{align}
	The purely imaginary eigenvalues of $B^\pm(\lambda)$ give us the Fredholm borders
	as curves in the complex plane parametrized by $\mu\in\mathbb{R}$
	\begin{align}
		\label{eq:grimshaw_fred}
		\lambda_1^\pm(\mu)&=iC\mu\notag\\
		\lambda_2^\pm(\mu)&=-i\Delta(K_z^\pm)\mu\notag\\
		\lambda_3^\pm(\mu)&=-\Lambda k^2(K_z^\pm)-\eta\hat{\omega}'(K_z^\pm)-i\Delta(K_z^\pm)\mu.
	\end{align}
	The Fredholm borders associated with $\lambda_1^\pm$ and $\lambda_2^\pm$ consist of the imaginary axis.
	They are invariant for every set of parameters defining 
	the wave solution and can not cause any unstable essential spectrum.

	However, the Fredholm borders associated with $\lambda_3^\pm$ have a real part 
	in which we can identify $p$ from \eqref{eq:grimshaw_p}.
	So, we substitute and see that
	\begin{align}
		\label{eq:grimshaw_rela}
		\Re(\lambda_3^\pm)&=\Delta(K_z^\pm)p(K_z^\pm)
	\end{align}
	and hence the stability of the wave packets depends on the sign of \eqref{eq:grimshaw_rela}. 
	In the previous section, we figured out that 
	\begin{align}
		\Delta<0\quad\text{for}\quad C>0,\\
		\Delta>0\quad\text{for}\quad C<0.
	\end{align}	
	Therefore, the sign of $\eqref{eq:grimshaw_rela}$ is determined by the sign of $p$ at the asymptotic rest states, 
	so the Fredholm border
	\begin{itemize}
		\item $\lambda_3^+$ is found on the right and $\lambda_3^-$ on the left hand side of the imaginary axis for $C>0$,
		\item $\lambda_3^-$ is found on the right and $\lambda_3^+$ on the left hand side of the imaginary axis for $C<0$.
	\end{itemize}

	To obtain the Fredholm indices in the delimited regions,
	we compute the Morse indices $i_\pm$ for real valued $\lambda\gg 1$,
	i.e. far to the right of the unstable $\lambda_3^\pm$.
	They are the sum of the algebraic multiplicities 
	of the unstable eigenvalues of the coefficient matrices $B^\pm(\lambda)$.
	For upward-traveling wave packets we find $i_\pm=3$ and for downward-traveling $i_\pm=0$.
	Thus, according to \eqref{eq:psinc_ind} the Fredholm index in this region is zero
	and the Fredholm indices of the remaining regions are calculated 
	in terms of the rule presented in section \ref{sec:psinc_stab} 
	for $\lambda$'s crossing the Fredholm borders.
	In particular, the region left of the rightmost Fredholm border $\lambda_3^\pm$
	has $\operatorname{ind}(\mathcal{L}_Y-\lambda)= +1$ and hence belongs to the essential spectrum 
	which is consequentially contained on the right hand side of the complex plane.
	To put it in a nutshell, the dissipative traveling wave packets are unconditionally unstable.

\begin{figure}[ht]
		\begin{center}
			\input{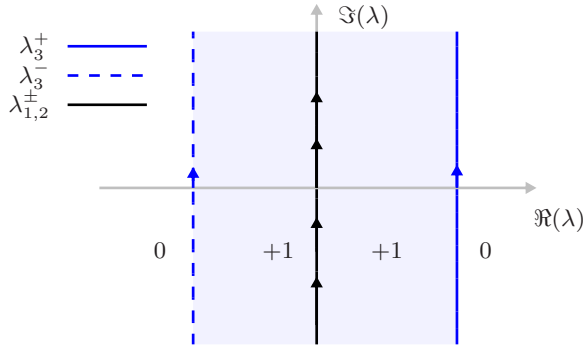}
		\end{center}
		\caption{Fredholm borders of $\mathcal{L}_Y$ for the dissipative pseudo-incompressible traveling wave packet and
		Fredholm indices.}
		\label{fig:psinc_dissipative_spec}
	\end{figure}

	The essential spectrum for the upward-traveling wave packet is plotted in figure \ref{fig:psinc_dissipative_spec}. 
	It consists of a vertical band in the complex plane delimited by the two Fredholm borders, 
	$\lambda_3^+$ to the right and $\lambda_3^-$ to the left.
	The remaining four borders lie on the imaginary axis. 
	For the downward-traveling wave packets the essential spectrum is very similar. 
	Only the orientation of the Fredholm borders switches and $\lambda_3^+$ swaps places with $\lambda_3^-$.
	The Fredholm indices stay untouched.
	
	We want to point out that \eqref{eq:grimshaw_rela} may give us a maximum instability growth rate
	which was a prerequisite for well-posedness as introduced in the beginning of this section.
	Therefore, it is worthwhile to investigate the spectrum and, by that, the nature of the instability in more detail.
	
	\subsection{Point spectrum and embedded eigenvalues}

	So far, we were only concerned with the essential spectrum,
	but to complete our argumentation about stability of the traveling wave packets 
	we have to take the point spectrum also into account (cf. section \ref{sec:psinc_stab}).
	We want to show that \eqref{eq:grimshaw_rela} yields indeed a maximum instability growth rate,
	i.e. there exists no point spectrum to the right of the essential spectrum.

	It suffices to consider the ODE associated with the eigenvalue problem for $\mathcal{L}_Y$ given by \eqref{eq:grimshaw_lin_op}.
	For upward-traveling waves 	in the limit $\zeta\rightarrow+\infty$ 
	the ODE has the coefficient matrix $B^+(\lambda)$ from \eqref{eq:grimshaw_bpm} 
	with eigenvalues $\nu^+$ having only positive real part for $\lambda$'s 
	in the regions to the right of the essential spectrum.
	With the help of theorem 11.2 in \citet[p 301]{Hartman2002} 
	using the eigenvalues $\nu^+$ one can show 
	that the corresponding ``eigenfunctions'' diverge at plus infinity and hence are not in $L^2$.
	For downward-traveling waves the same argument holds at minus infinity.
	In conclusion, for both traveling wave solutions there is no unstable point spectrum
	and the rightmost Fredholm border gives a bound on the instability growth rate.
	Hence, the system was regularized in the desired sense by including dissipative effects.
	
	In the remainder of this section we want to discuss how positive real valued $\lambda$'s in the spectrum
	manifest themselves as instabilities.
	We observe that the Fredholm index in the unstable essential spectrum is $+1$ for both upward and downward-traveling wave packets.
	By the definition \eqref{eq:psinc_def_fred},
	a positive Fredholm index implies that the kernel of $\mathcal{L}_Y-\lambda$ has at least dimension one
	and is in particular non-empty.
	Hence, there exist $y\in L^2, y\neq 0$ that solve $\mathcal{L}_Y\,y=\lambda\,y$.
	Those $y$'s can be called eigenfunctions and the corresponding $\lambda$'s eigenvalues.
	In conclusion, the unstable essential spectrum consists of embedded eigenvalues
	and the instabilities are superpositions of exponentially growing eigenfunctions.

	This observation is important as it prohibits convective instabilities \citep{Sandstede2000}
	which are characterized by an unstable essential spectrum that can be stabilized in an 
	exponentially weighted space (instead of the usual $L^2$). 
	Convective instabilities are transported on the domain towards the infinities away from the traveling wave solution
	such that the solution can survive.
	Even though it is possible here to stabilize the essential spectrum in our case,
	the interpretation that the instability may be convective is flawed due to the embedded eigenvalues.

	\subsection{Numerical simulation of the dissipative traveling wave packet}
	\label{sec:grimshaw_num}
	
	In this section we will analyze the essential instabilities 
	of the dissipative traveling wave packets numerically.
	The modulation equations \eqref{eq:grimshaw_mod} are solved with the method presented in appendix A.
	Let us investigate the two classes of dissipative traveling wave solutions 
	separately starting with the upward-traveling wave packet.
	The parameters defining the wave are given by $N=2.1$, $K_x=1.5$, $\Lambda=1.1$ and $C=1.6$\,.
	The initial conditions are computed by the fourth-order Runge-Kutta method 
	solving the ODE \eqref{eq:grimshaw_ode3}.
%	%
	\begin{figure}[ht]
		\centering
		\input{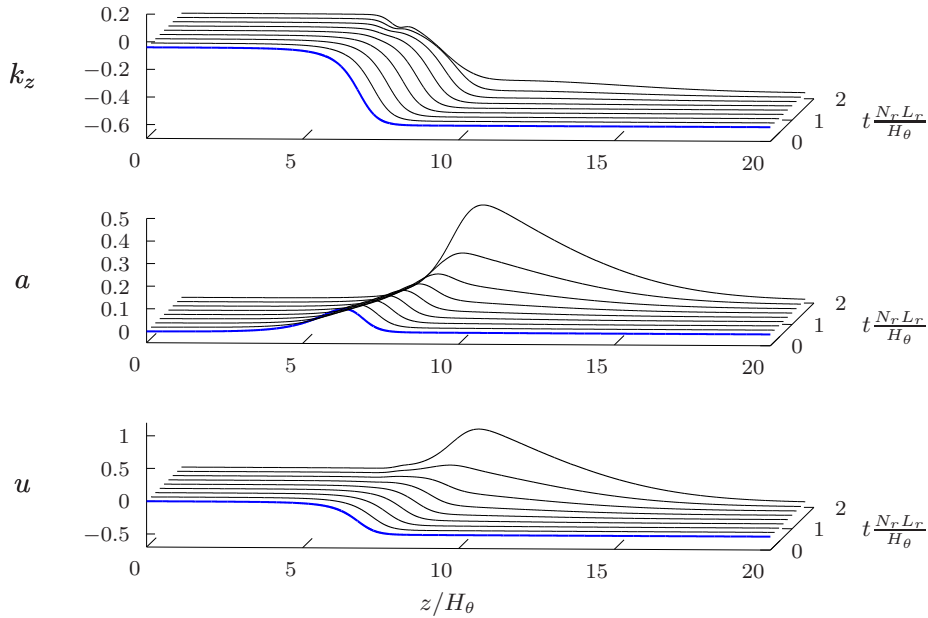}
		\caption{Numerical simulation of the dissipative Grimshaw upward-traveling wave packet experiencing transient instability.}
		\label{fig:grimshaw_upward}
	\end{figure}
	The results of the numerical time integration are shown in figure \ref{fig:grimshaw_upward}. 
	The labeling of axes is the same as in section \ref{sec:psinc_num} 
	for the numerical simulations of the inviscid waves.
	An instability is excited by an initial perturbation of Gaussian shape of the order of magnitude $10^{-3}$ added to the wave packet.
	The instability growth quickly but,
	in contrast to the two inviscid cases of the previous sections,
	it possesses a large scale structure.
	In terms of \eqref{eq:grimshaw_rela} we can compute the maximum instability growth rate analytically 
	and find $\Re(\lambda)_\mathrm{max}=5.1$\,.
	By computing the $L^2$-norm of the difference between the numerical and a reference solution,
	the growth rate of the actually excited instability can be computed numerically.
	It is approximately 3.8\,. This rate is smaller than the expected value 
	because the initial perturbation was not optimal and the numerical error is of diffusive character.
	
%	We can observe that the instability arisen at the tail 
%	of the packet detaches from the wave in agreement with the analytical findings of the previous section.
%	It travels downwards, i.e. towards $-\infty$, in the moving frame 
%	being identical with an immobile instability in the fixed frame of reference.
%	Remarkably, the transient instability grows in the vertical wave number 
%	whereas it decays in the specific wave action density and the mean-flow horizontal wind.
%	The time domain being displayed converts in dimensional units 
%	to several hours for reference values typical for the mesosphere.

\begin{figure}[ht]
		\centering
		\input{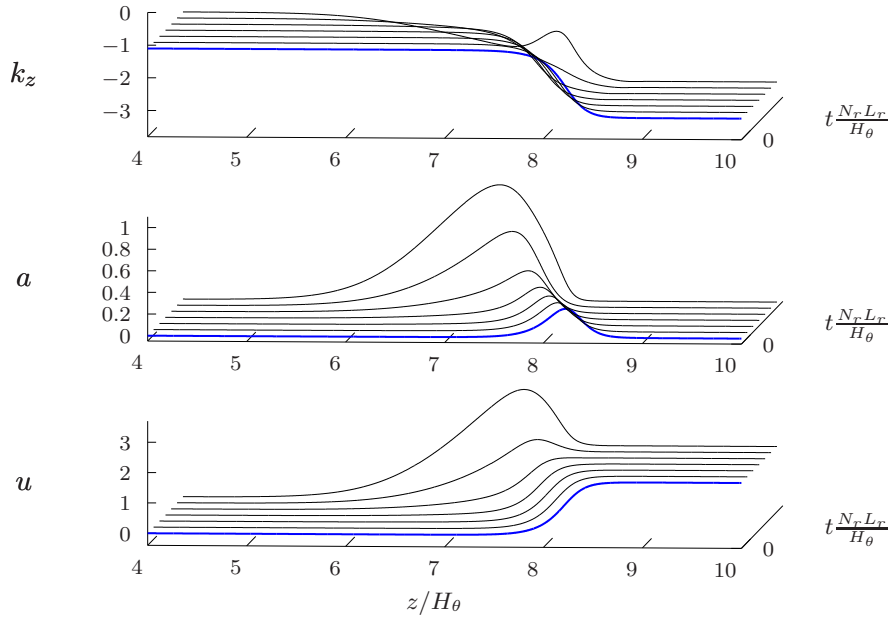}
		\caption{Numerical simulation of the dissipative Grimshaw
		downward-traveling wave packet showing transient instability. 
		}
		\label{fig:grimshaw_downward}
	\end{figure}

	Let us continue with the dissipative downward-traveling wave packet being 
	defined through the parameters $N=2.2$, $K_x=1.3$, $\Lambda=1.1$ and $C=-0.7$\,.
	Likewise, it is initially perturbed by a Gaussian $10^{-3}$-deviation from the exact traveling wave solution.
	We plotted the results of the numerical time integration in figure \ref{fig:grimshaw_downward}
	where we can observe that the wave blows up by a large-scale instability.
	The maximum growth rate according to \eqref{eq:grimshaw_rela} is 8.9\,.
	And the approximated numerical growth rate of the actually excited instability is 6.8\,.
	In conclusion, the numerical results fit nicely into our theoretical framework from the spectral stability analysis.

%	%
%	Numerical results are depicted in figure \ref{fig:grimshaw_downward}.
%	This wave packet features not only a very large amplitude but also an extreme narrow extent.
%	However, the wave is able to sustain for a long time 
%	after the transient instability detaches from the packet.
%	The instability is only noticeably visible in the vertical wave number
%	and slightly recognizable in the mean-flow horizontal wind in the tail of the wave
%	where it propagates in the moving frame towards $+\infty$ as expected from the stability analysis.
	Note that the linear group velocity of the downward-traveling wave packet is positive everywhere 
	whereas the envelope velocity is negative. 
	This seemingly paradoxical situation is a logical consequence of the nonlinear nature of these waves.

	\section{Conclusion}
	\label{sec:conclusion}
	
%%% Results
	We studied the modulation equations of two-dimensional 
	nonlinear gravity waves in the isothermal atmosphere and their stability.
	The waves were assumed to be horizontally homogeneous.
	Two different asymptotic regimes of the Euler equations were of interest, 
	the Boussinesq and the pseudo-incompressible equations.
	The former does not account for altitudinal variations	of the background density 
	but possesses the convenient property of having plane wave solutions
	that we examined with respect to stability introducing the notion of spectral stability analysis.
	Here, we were able to reproduce the criterion which 
	is known from modulational instability of the weakly nonlinear theory.
	Allowing waves to propagate on times scales where the background density changes,
	brought us to the inviscid pseudo-incompressible modulation equations.
	They exhibit solutions of the form of upward-traveling wave fronts
	which destabilize unconditionally.

	The unstable operator spectra of the linearized inviscid 
	modulation equations of the two regimes were revealed to be unphysical.
	Mathematically speaking, the systems were ill-posed.
	Therefore, we deduced that they may be regularized 
	by considering the compressible Navier-Stokes instead of the Euler equations.
	We found that the emerging dissipative modulation equations can have
	upward and downward-traveling wave packet solutions.
	The spectral stability analysis showed that the wave packets 
	are unstable due to embedded eigenvalues.
	
	We emphasized that linear group and envelope velocities can differ for nonlinear waves.
	Remarkably, the dissipative  downward-traveling wave packet is an especially interesting case.
	Its phase and envelope velocities are negative, 
	though the linear group velocity is positive everywhere.

\bibliography{library}
%\printbibliography[heading=bibintoc]

\end{document}